\documentclass[runningheads]{llncs}
\usepackage[T1]{fontenc}
\usepackage{graphicx,verbatim}
\usepackage{amssymb}
\usepackage{latexsym}
\usepackage{amsmath}
\usepackage{algorithm}
\usepackage{algpseudocode}
\usepackage{graphicx}
\usepackage{textcomp}
\usepackage{booktabs}
\usepackage{color}
\usepackage{multirow}
\usepackage{array}
\usepackage{multicol}
\usepackage{layouts}
\usepackage{braket}
\usepackage{siunitx}
\usepackage{subfigure}
\usepackage{float}
\usepackage{hyperref}
\usepackage{bm}
\begin{document}
%
%\title{Accelerated Cardiac $T_1$ Mapping with Physics-Informed Neural Networks}
\title{Physics-Informed Neural ODEs for Temporal Dynamics Modeling in Cardiac $T_1$ Mapping}
\author{Nuno Capitão \inst{1,2}\textsuperscript{*} % index{Capitão, Nuno}
\and Yi Zhang\inst{1}\textsuperscript{*} % index{Zhang, Yi}
\and
Yidong Zhao \inst{1} % index{Zhao, Yidong}
\and
Qian Tao \inst{1} % index{Tao, Qian}
}
\authorrunning{N. Capitão et al.}
\institute{Department of Imaging Physics, Delft University of Technology, the Netherlands \and University of Porto, Portugal \\\textsuperscript{*} These authors contributed equally to this work. \\ \email{\{up201709363\}@up.pt, \{y.zhang-43,y.zhao-8,q.tao\}@tudelft.nl}}

\maketitle              % typeset the header of the contribution
\begin{abstract}
Spin-lattice relaxation time ($T_1$) is an important biomarker in cardiac parametric mapping for characterizing myocardial tissue and diagnosing cardiomyopathies. Conventional Modified Look-Locker Inversion Recovery (MOLLI) acquires 11 breath-hold baseline images with interleaved rest periods to ensure mapping accuracy. However, prolonged scanning can be challenging for patients with poor breathholds, often leading to motion artifacts that degrade image quality. In addition, $T_1$ mapping requires voxel-wise nonlinear fitting to a signal recovery model involving an iterative estimation process. Recent studies have proposed deep-learning approaches for rapid $T_1$ mapping using shortened sequences to reduce acquisition time for patient comfort. Nevertheless, existing methods overlook important physics constraints, limiting interpretability and generalization. In this work, we present an accelerated, end-to-end $T_1$ mapping framework leveraging Physics-Informed Neural Ordinary Differential Equations (ODEs) to model temporal dynamics and address these challenges. Our method achieves high-accuracy $T_1$ estimation from a sparse subset of baseline images and ensures efficient null index estimation at test time. Specifically, we develop a continuous-time LSTM-ODE model to enable selective Look-Locker (LL) data acquisition with arbitrary time lags. Experimental results show superior performance in $T_1$ estimation for both native and post-contrast sequences and demonstrate the strong benefit of our physics-based formulation over direct data-driven $T_1$ priors.

\begin{figure}[!htbp]
\includegraphics[width=\textwidth]{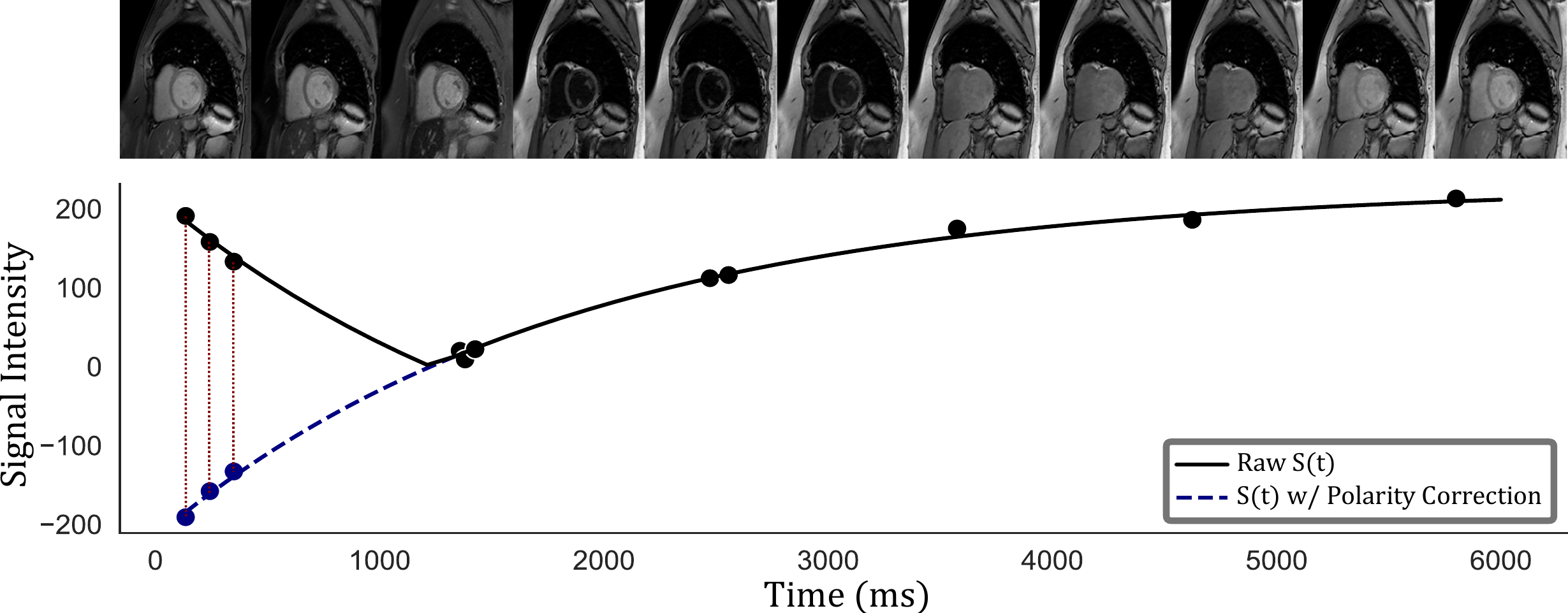}
\caption{MOLLI 3(3)3(3)5 reconstruction scheme illustrating signal recovery with and without polarity recovery. The corresponding temporally ordered MOLLI $T_1$-weighted images are shown above.
} 
\label{fig: molli_reconstruction}
\end{figure}

% Authors must provide keywords and are not allowed to remove this Keyword section.

\end{abstract}
\section{Introduction}
Quantitative Magnetic Resonance Imaging (qMRI) is a widely adopted noninvasive technique for cardiac parametric mapping, where spin-lattice relaxation time ($T_1$) is routinely used as a biomarker to clinically assess myocardial tissue and diagnose cardiomyopathies \cite{clinicalT1}. Cardiac $T_1$ maps are generated by fitting a parametric model to a series of baseline images that encode signal recovery governed by spin-lattice energy exchange after an inversion pulse.

The Modified Look-Locker Inversion Recovery (MOLLI) \cite{MOLLI} sequence has become the standard protocol in cardiac qMRI due to its high precision and reduced motion-induced artifacts. During a single breath hold, MOLLI acquires $T_1$-weighted images at multiple inversion times using distinct Look-Locker (LL) experiments interleaved with rest periods. In a MOLLI 3(3)3(3)5 sequence, three LL experiments collect images in sets of 3, 3, and 5, each separated by three heartbeats. These acquisitions form a image series of 11 baseline images that span the full relaxation curve, as illustrated in Figure \ref{fig: molli_reconstruction}. While multiple LL experiments at different inversion times improve $T_1$ estimation accuracy, they also extend the breath-hold to 17 heartbeats, which can be strenuous for patients with respiratory issues. Alternative schemes such as Shortened MOLLI (ShMOLLI) \cite{ShMOLLI} and Saturation Recovery Single-Shot Acquisition (SASHA) \cite{SASHA} reduce the total number of acquisitions to accelerate imaging, but inherently trade off precision and accuracy \cite{WEINGARTNER1,WEINGARTNER2}.

Traditional $T_1$ map estimation requires voxel-wise nonlinear curve fitting to a signal recovery model derived from MRI physics. This process is further complicated by the need to identify signal polarity transitions (null points) in MOLLI sequences, requiring additional iterative calculations for each pixel. The entire procedure is not only sensitive to image noise but also computationally intensive - processing a complete 3D volume over time has $\mathcal{O}(n^4)$ complexity, resulting in significant processing delays in clinical settings without specialized acceleration techniques \cite{zhang2024torcht}. 

Recent studies have shown that deep learning methods can accurately and efficiently perform cardiac $T_1$ mapping using as few as three to five baseline images along the inversion recovery curve, achieving results comparable to conventional MOLLI. \cite{MyoMapNet} introduced MyoMapNet, a fast myocardial $T_1$ mapping model that uses a fully connected neural network (FCNN) to directly map pairwise LL readouts to $T_1$ values. However, it neglects signal polarities and is constrained to a data-driven $T_1$ prior. In contrast, T1Net \cite{T1Net} employs a bidirectional long short-term memory (LSTM) network \cite{LSTM} and adheres to the signal recovery model to enforce consistency between acquired MRI signals and predicted $T_1$ values through its loss function. Nonetheless, we hypothesize that T1Net remains limited for several reasons: First, estimating $T_1$ maps from a single LL experiment may require non-uniform time intervals to capture the full relaxation curve. Like other discrete recurrent neural networks, LSTMs struggle with irregular time gaps \cite{NCDE,schirmer2022modelingirregulartimeseries,T-LSTM}. Furthermore, although enforcing the signal recovery model provides a stronger prior than direct $T_1$ estimation, the mechanism by which the model learns the relaxation rate remains a black box, limiting both interpretability and generalization.

These limitations highlight the need for a more principled approach that can handle irregular sampling patterns while maintaining physical consistency with the underlying MR signal dynamics. To address these challenges, we introduce a robust and accelerated $T_1$ mapping technique using physics-informed neural networks (PINNs) \cite{PINN}. We employ a continuous-time LSTM-ODE \cite{LSTMODE} that can naturally handle LL sequences with non-uniform acquisition intervals, overcoming the time-gap limitations of traditional recurrent networks. Our Neural Ordinary Differential Equation (NODE) \cite{NODE} approach explicitly models the temporal dynamics of the relaxation process, providing a continuous representation of signal evolution rather than discrete approximations. This dynamic modeling not only adheres to the signal recovery model but also explicitly formulates the recovery rate along the relaxation trajectory, ensuring that $T_1$ estimation follows the actual physics of spin-lattice interactions. Furthermore, the framework seamlessly incorporates signal polarity correction during inference, eliminating the need for separate processing steps. %This integrated approach significantly reduces both acquisition time and computational complexity, potentially enabling wider clinical adoption of cardiac $T_1$ mapping, particularly for patients with limited breath-hold capabilities.

\section{Methods}
\subsection{MR Physics-Informed Signal Modeling and Loss Functions}

To develop our physics-informed approach, we first formalize the MR signal model that governs $T_1$ relaxation. In qMRI, $N$ baseline images are collected via selective sampling, with each image capturing the signal at a different point along the relaxation curve. The signal intensity $S(x,y,z)$ at a given spatial coordinate $(x,y,z)$ is described by a parametric function that reflects the underlying physics. In a MOLLI sequence, the signal recovery follows a 3-parameter model expressed as:

\begin{equation}
\label{eq: signal_recovery_model}
S_{(x,y,z)}(t_i) = c_{(x,y,z)} \cdot (1 - k_{(x,y,z)}\cdot \exp(-\frac{t_i}{T_{1,(x,y,z)}^{*}})),
\end{equation}
where $i \in \{1,2,\dots, N\}$ and $t_i$ is the inversion time of the $i$-th image. The parameter set $\{c, k, T_{1}^{*}\}$ is estimated at each voxel $v \in \mathbb{R}^{x \times y \times z}$ to determine $T_1$ biomarker using the following equation:

\begin{equation}
\label{eq: T1_equation}
T_{1 (x,y,z)} = T_{1,(x,y,z)}^{*}\cdot(k_{(x,y,z)} - 1).
\end{equation}
While Equations \eqref{eq: signal_recovery_model} and \eqref{eq: T1_equation} provide the foundation for conventional $T_1$ mapping, they represent a static view of the relaxation process. Our approach incorporates the underlying physical dynamics governing spin-lattice relaxation by explicitly modeling the temporal evolution of the signal. By differentiating Equation \ref{eq: signal_recovery_model}, we derive a first-order differential expression that captures the instantaneous rate of change at each inversion time $t_i$:

\begin{equation}
\label{eq: signal_recovery_derivative}
\frac{dS_{(x,y,z)}(t_i)}{dt} = \frac{c_{(x,y,z)} \cdot k_{(x,y,z)}}{T_{1,(x,y,z)}^{*}}\cdot  \exp\left(-\frac{t_i}{T_{1,(x,y,z)}^{*}}\right).
\end{equation}
This differential formulation enables our model to learn not just the signal values but also their temporal dynamics, providing a more complete physical representation of the relaxation process.

To incorporate both data-driven priors and physics-based constraints into our deep-learning framework, we define a physics-informed loss function with two components: a $T_1$ consistency loss $\mathcal{L}_{T_1}$ and a physics-based loss $\mathcal{L}_{\text{physics}}$.
\begin{equation}
\label{eq: loss_data}
\mathcal{L}_{T_1} = \frac{1}{HWD} \sum_{(x,y,z)} \|T_1 - \hat{T_1}\|_2^2,
\end{equation}
where $T_1$ is derived from Equation \eqref{eq: T1_equation}, enforcing consistency with our end-goal parametric mapping.
\begin{equation}
\label{eq: loss_physics}
\mathcal{L}_{\text{physics}} = \frac{1}{NHWD} \sum_{(x,y,z)} \sum_{i=1}^{N} \|S(t_i) - \hat{S}(t_i)\|_2^2  + \lambda \|\gamma\frac{dS(t_i)}{dt} - \frac{d\hat{S}(t_i)}{dt}\|_2^2.
\end{equation}
The first term ensures adherence to the signal recovery model (Equation \eqref{eq: signal_recovery_model}), while the second term penalizes deviations from the expected signal dynamics (Equation \eqref{eq: signal_recovery_derivative}). The scaling factor $\lambda$ adjusts for the disparity in orders of magnitude observed at $\lim_{t \to 0} \tfrac{dS(t)}{dt}$ relative to the other loss terms. The factor $\gamma$ is derived from the chain rule and accounts for the normalization ratio applied to signals and inversion times. 
\begin{figure}[!ht]
\includegraphics[width=\textwidth]{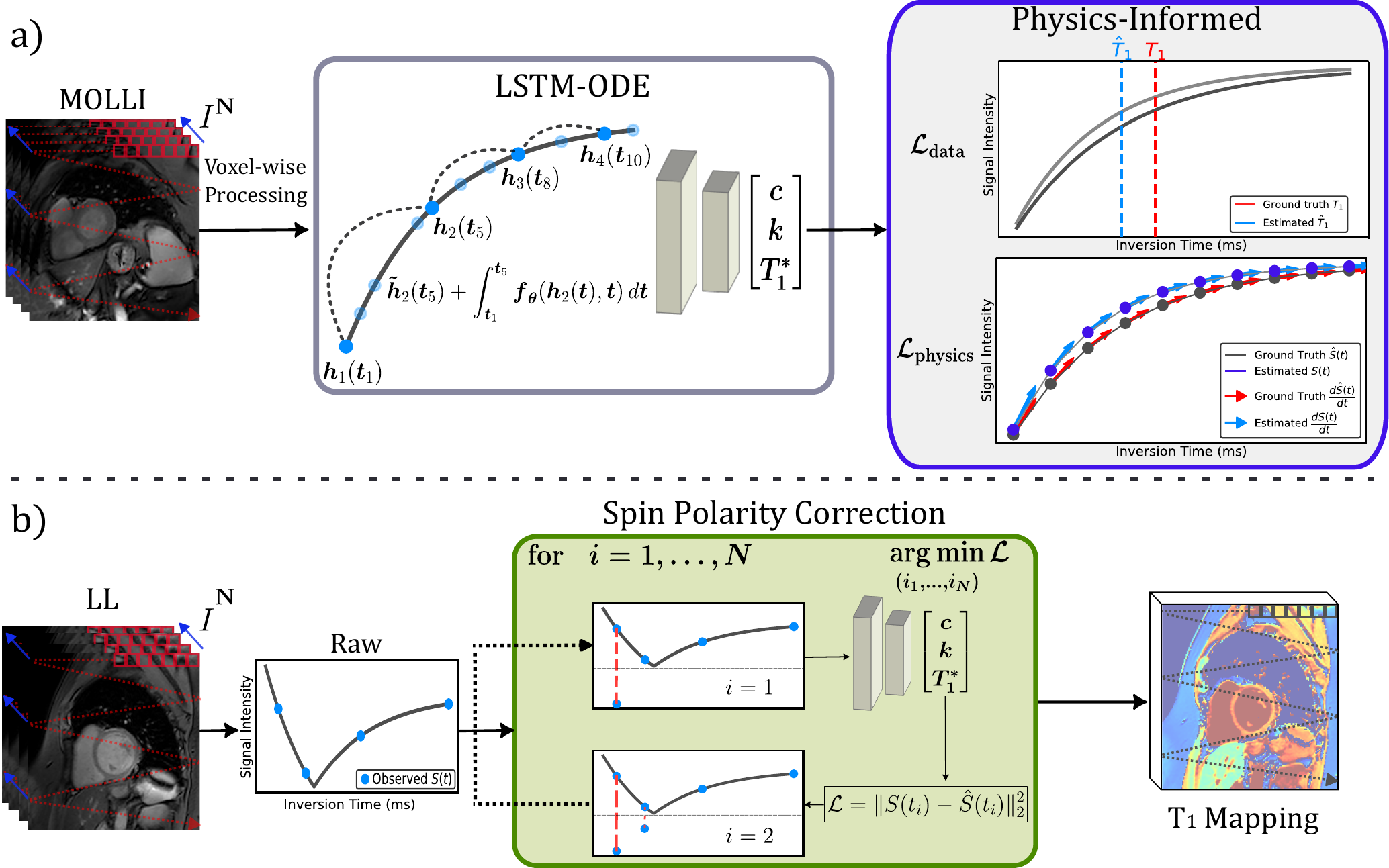}
\caption{Overview of the proposed physics-informed LSTM-ODE framework (a) \textbf{Training:} The model processes a subset of the 11 MOLLI baseline images in a voxel-wise manner through a continuous-time LSTM-ODE. The physics-informed loss is computed using $\{c, k, T_1^*\}$, estimated by the model. (b) \textbf{Inference:} For a series of $I^{N}$ LL images, each voxel undergoes spin polarity correction. Every $i$-th image is sequentially inverted, and the resulting sequence is evaluated using the signal reconstruction loss. The index that minimizes the loss, $\arg\min \mathcal{L}$, determines the null index and the optimal parameters $\{c, k, T_1^*\}$.
} 
\label{fig: PINN_LSTM_ODE}
\end{figure}
By combining these complementary loss components, we formulate our final loss function as:
\begin{equation}
\label{eq: total_loss}
\mathcal{L}_{\text{total}} = \mathcal{L}_{T_1} + \mathcal{L}_{\text{physics}}.
\end{equation}

\subsection{Continuous-Time Neural ODEs for Temporal Dynamics Modeling}

To effectively capture the temporal dynamics of spin-lattice relaxation with non-uniform sampling intervals, we implement a continuous-time LSTM-ODE framework that naturally handles the irregular time gaps in MOLLI sequences. Let $\{(S_i, t_i)\}_{i=1}^N$ represent a MOLLI voxel sequence, where $S_i$ is the measured signal at inversion time $t_i$. At the $i$‑th image, the LSTM's discrete gating update is determined by
\begin{equation}
    (\tilde{h}_i,\, c_i) = \mathrm{LSTMCELL}\bigl((S_i,\, t_i),\,(h_{i-1},\, c_{i-1})\bigr),
\end{equation}
where $h_{i-1}$ and $c_{i-1}$ are the previous hidden and cell states, respectively, $\tilde{h}_i$ denotes the candidate hidden state for the current image, and $c_i$ represents the updated cell state. The candidate hidden state $\tilde{h}_i$ is then evolved continuously over the interval $[t_{i-1}, t_i]$ by a NODE whose dynamics are governed by
\begin{equation}
    \frac{d\,h(t)}{dt} = f_{\theta}\bigl(h(t),\, t\bigr),
\end{equation}
where $f_{\theta}$ is a small, learnable neural network. We initialize the ODE at $h(t_{i-1}) = \tilde{h}_i$ and integrate forward using a black-box 5th‑order Dormand–Prince solver \cite{KONSTANTINOV198027}. The cell states $\{c_i\}_{i=1}^N$ remain discrete to preserve stable long-term gradient propagation \cite{LSTMODE}. The final hidden state, $h_N(t_N)$, provides a continuous‑time encoding of the relaxation trajectory, capturing the temporal variations in signal recovery across the $N$ images. A multi-layer decoder then maps this encoding to the signal recovery model described by Equations (\ref{eq: signal_recovery_model}–\ref{eq: signal_recovery_derivative}). Formally, the mapping notation can be defined as $f_{\theta}: \{(S_i, t_i)\}_{i=1}^N \in \mathbb{R}^{N \times d_\text{emb} \times 2} \to \mathbb{R}^{3}$, where $d_\text{emb}$ denotes the feature dimension of the embedded pairwise acquisitions, and the output represents the set of parameters $\{C, k, T_1^*\}$.

This continuous-time modeling approach enables our framework to accurately represent the underlying physics of relaxation processes while accommodating the practical constraints of clinical MRI acquisition protocols with varying inversion times.

\subsection{Post-Inversion Spin Polarity Correction}
At test time, raw signal acquisitions retain their absolute magnitudes, which conflicts with our model's assumptions about post-inversion signal recovery. However, since our method is formulated around the signal recovery model, it can also act as a proxy for retrieving the spin state polarity encoded in each image. To estimate the null index at each voxel, we evaluate all possible polarity permutations across the sequence by performing $N$ forward passes, leveraging vectorized, batch-wise inference over the full field-of-view (FoV). Figure \ref{fig: PINN_LSTM_ODE} illustrates our end-to-end $T_1$ parametric mapping framework, which includes the post-inversion spin polarity correction. During the polarity recovery step, each $i$-th signal is inverted, and the entire sequence undergoes a model evaluation to compute the signal reconstruction loss term in Equation \ref{eq: loss_physics}. For each voxel $v \in \mathbb{R}^{x \times y \times z}$, the $\arg\min \mathcal{L}$ across trials identifies the null index, providing the best estimate of $\{c, k, T_1^*\}$ for $T_1$ mapping.

\section{Experiments}
\textbf{Dataset:} We used a cardiac MRI dataset of 50 subjects acquired with a 3.0T Ingenia MR scanner (Philips Healthcare). Each subject underwent both pre- and post-contrast MOLLI sequences, each consisting of 11 baseline images using the same 3-3-5 scheme provided by the manufacturer. One to three axial slices were acquired at the base, mid-ventricular, and apical levels, and the left-ventricle myocardium was manually annotated as the region of interest (ROI). To prevent data leakage, we performed a subject-wise random split: 30 subjects for training, 5 for validation, and 15 for testing. To evaluate out-of-domain generalization in signal recovery patterns at higher relaxation rates induced by a contrast agent, the training set included only native sequences, while the validation and testing sets contained both pre- and post-contrast sequences. \newline

\noindent \textbf{Implementation Details:} We pre-train our LSTM-ODE model on all $11$ MOLLI images and fine-tune on subsets of $3$-$5$ images to simulate accelerated acquisition. During training, sparse readouts are strategically selected to ensure coverage across the relaxation curve: early phase ($\{t_1$-$t_3\}$), intermediate phase ($\{t_4$-$t_8\}$), and convergence phase ($\{t_9$-$t_{11}\}$). All models are trained using ground-truth $\{c, k, T_1^*\}$ parameters fitted to the $11$-image data with the Levenberg-Marquardt (LM) algorithm. For the physics-based loss, we generate $1\,000$ interpolated points with $75\%$ linearly spaced up to $2000$\,ms to emphasize relaxation dynamics. Our ODE solver uses tolerances of $\epsilon = 0.001$ and runs on an NVIDIA GTX $1070$ GPU. We employ standard backpropagation with a derivative loss hyperparameter of $\lambda = 0.01$. Our implementation will be released on GitHub. \newline

\noindent \textbf{Comparative Studies:}
We benchmark our baseline model against three methods: MyoMapNet, T1Net (both designed for accelerated $T_1$ mapping), and SciPy's Trust Region Reflective (TRF) algorithm for bounded parameter constraints. To evaluate whether our physics-based priors improve performance relative to direct data-driven $T_1$ priors, we also integrate MyoMapNet's architecture into our framework for direct comparison. \newline

\noindent \textbf{Evaluation Metrics:}
We perform 100 Monte Carlo simulations per subject using random sampling to assess generalization across LL subsets. We evaluate two metrics: (1) mean $T_1$ bias relative to ground truth, and (2) fitting standard deviation (SD), which measures curve fitting quality at each voxel from the fitting residuals~\cite{kellman2013t1,kellman2014t1,zhang2024deep}. Both metrics are calculated within myocardium regions annotated by radiologists. Pixel-wise $T_1$ estimates are averaged over simulations to generate subject-level distributions, and systematic biases are evaluated using paired $t$-tests.

\section{Results and Discussion}
\begin{table}[t]
\centering
\fontsize{8pt}{9.6pt}\selectfont
\caption{Mean $T_1$ bias and SD values from Monte Carlo simulations, expressed as mean(std). Methods: TRF, MMNet (MyoMapNet), T1N (T1Net), P-MM (PINN MyoMapNet), P-LSTM (PINN LSTM-ODE). The best-performing methods are in \textbf{bold} and second-best are \underline{underlined}. For bias values, $^*$ indicates no statistical significance ($p > 0.05$).}
\label{table: combined_results}
\begin{tabular}{lccccccc}
\toprule
\multirow{2}{*}{Method} & \multicolumn{3}{c}{Mean Bias (ms)} & \multicolumn{3}{c}{Fitting SD (ms)} \\ 
\cmidrule(lr){2-4} \cmidrule(lr){5-7}
& $\text{LL}_3$ & $\text{LL}_4$ & $\text{LL}_5$ & $\text{LL}_3$ & $\text{LL}_4$ & $\text{LL}_5$ \\ 
\midrule
\multicolumn{7}{l}{\textbf{Native}} \\
\midrule 
TRF & -9.63(38.21) & -12.90(31.25) & 15.64(18.80) & \textbf{58.12}(37.87) & \textbf{49.66}(27.16) & \textbf{46.83}(24.12) \\
MMNet & 51.50(11.44) & 21.49(10.13) & 11.29(9.26) & 91.57(36.67) & 67.49(26.94) & \underline{50.47}(24.36) \\
T1N & \underline{-11.28}(8.83) & \underline{-8.77}(11.75) & \underline{3.11}$^*$(9.74) & 94.00(34.81) & 69.17(29.87) & 53.33(26.82) \\
P-MM & \textbf{0.15}$^*$(6.62) & 5.39$^*$(11.70) & 7.71(8.90) & \underline{82.74}(37.10) & \underline{62.40}(31.39) & 51.19(27.60) \\
P-LSTM & 7.65(9.93) & \textbf{2.62}$^*$(7.74) & \textbf{0.75}$^*$(8.68) & 91.18(38.74) & 70.04(30.09) & 50.65(27.43) \\
\midrule
\multicolumn{7}{l}{\textbf{Post-Gd}} \\
\midrule
TRF & 16.68(24.57) & 7.64(14.57) & 11.83(12.65) & \textbf{46.49}(37.58) & \textbf{39.95}(17.48) & \underline{43.27}(16.95) \\
MMNet & \underline{1.18}(22.29) & 14.23(13.86) & 11.40(11.58) & \underline{73.67}(17.32) & \underline{58.34}(15.55) & 42.65(20.45) \\
T1N & \textbf{8.75}(10.62) & \underline{8.75}(10.42) & 20.44(7.82) & 79.99(21.78) & 53.97(17.18) & 41.00(16.05) \\
P-MM & -6.52(18.47) & 5.94(11.82) & \underline{11.83}(9.65) & 86.93(22.67) & 59.24(18.94) & \textbf{40.29}(16.93) \\
P-LSTM & -12.37(18.12) & \textbf{-1.12}$^*$(11.48) & \textbf{3.38}$^*$(11.11) & 110.75(52.40) & 70.55(18.75) & 42.92(20.47) \\
\bottomrule
\end{tabular}
\end{table}

%\begin{figure}[!htbp]
%\centering
%\includegraphics[width=0.8\textwidth]{figures/results_postgd_smaller.pdf}
%\caption{results_postgd
%} 
%\label{fig: results_postgd}
%\end{figure}

\begin{figure}[!htb]
    \centering
    \begin{tabular}{cc}
        \includegraphics[width=0.5\textwidth]{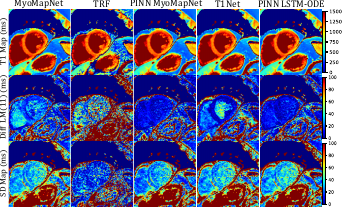} &
        \includegraphics[width=0.5\textwidth]{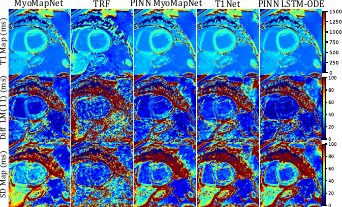} \\
    \end{tabular}
    \caption{Estimated mean values of $T_1$ and SD maps. Left: native pre-contrast sequence. Right: post-contrast sequence.}
    \label{fig: qualitative_results}
    \end{figure}

Table \ref{table: combined_results} presents our Monte Carlo simulation results, summarizing the mean $T_1$ bias and SD distributions for Native and Post-Gd testing sets. Complementary qualitative comparisons for cardiac $T_1$ mapping are provided in Figure \ref{fig: qualitative_results}. The PINN MyoMapNet model improved upon conventional MyoMapNet, demonstrating the benefits of incorporating physics during training. Our PINN LSTM-ODE method achieved the overall lowest mean bias for both Native and Post-Gd settings, indicating stronger generalization. However, its higher SD maps suggest that sparse representations may lead to inconsistent NODE-computed trajectories. The TRF algorithm performs subpar in terms of mean bias, highlighting the potential of deep learning for higher accuracy in $T_1$ mapping given a limited number of baseline images. Nevertheless, TRF reports the lowest SD map statistics, suggesting that deep learning methods still compromise precision compared to traditional nonlinear curve-fitting approaches.
\section{Conclusions}
We proposed a novel deep learning framework to accelerate MOLLI-based cardiac $T_1$ mapping by integrating PINNs. Our experimental results show that the proposed PINN LSTM-ODE model outperforms alternative methods in both accuracy and precision, particularly with limited LL samples. Furthermore, we demonstrated that physics-informed formulations improve the performance of models that traditionally rely on direct data-driven $T_1$ priors, such as MyoMapNet. This highlights the benefits of embedding physics-based constraints into model training. The proposed framework enables accurate $T_1$ estimation and efficient FoV mapping from just 3–5 heartbeats, matching the performance of the conventional 17-heartbeat MOLLI sequence while significantly reducing acquisition time.

\begin{comment}  %% removed for anonymized MICCAI 2025 submission.
    
    % The following acknowledgement and disclaimer sections should be removed for the double-blind review process.  
    % If and when your paper is accepted, reinsert the acknowledgement and the disclaimer clause in your final camera-ready version.

\begin{credits}
\subsubsection{\ackname} A bold run-in heading in small font size at the end of the paper is
used for general acknowledgments, for example: This study was funded
by X (grant number Y).

\subsubsection{\discintname}
It is now necessary to declare any competing interests or to specifically
state that the authors have no competing interests. Please place the
statement with a bold run-in heading in small font size beneath the
(optional) acknowledgments\footnote{If EquinOCS, our proceedings submission
system, is used, then the disclaimer can be provided directly in the system.},
for example: The authors have no competing interests to declare that are
relevant to the content of this article. Or: Author A has received research
grants from Company W. Author B has received a speaker honorarium from
Company X and owns stock in Company Y. Author C is a member of committee Z.
\end{credits}

\end{comment}
\begin{credits}
\subsubsection{\ackname} We gratefully acknowledge the Dutch Research Council (NWO) and Amazon Science for financial and computing support.

\subsubsection{\discintname}
The authors have no competing interests to declare that are
relevant to the content of this article.
\end{credits}
\bibliographystyle{splncs04}
\bibliography{paper-4475}
%\bibitem{ShMOLLI}
%Piechnik, S.K., Ferreira, V.M., Dall'Armellina, E. et al. Shortened %Modified Look-Locker Inversion recovery (ShMOLLI) for clinical %myocardial T1-mapping at 1.5 and 3 T within a 9 heartbeat %breathhold. \textit{J Cardiovasc Magn Reson} \textbf{12}, 69 (2010). %\doi{10.1186/1532-429X-12-69}

%\end{thebibliography}
\end{document}